\documentclass[cameraready]{Interspeech}
\title{Ouroboros: Self-Referential Backdoor Attacks on Speech Enhancement via Clean Audio Triggers}

\author[affiliation={1}]{Yunjie}{Zhou}
\author[affiliation={1}]{Yuheng}{Huang}
\author[affiliation={1,2}]{Diqun}{Yan}
\address{
    $^1$ Faculty of Electrical Engineering and Computer Science, Ningbo University, China \\
    $^2$ College of Artificial Intelligence, Ningbo University of Finance and Economics, China 
}

\email{2411100321@nbu.edu.cn, 2311100082@nbu.edu.cn, yandiqun@nbufe.edu.cn}


\usepackage{comment}

\usepackage{graphicx}
\usepackage{amsmath,amssymb,amsfonts}
\usepackage{cleveref}
\usepackage{multirow}
\usepackage{booktabs}
\usepackage{threeparttable}
\usepackage{flushend}
\usepackage{tikz}

\usepackage{booktabs} % 三线表
\usepackage{multirow} % 跨行表格
\usepackage{arydshln} % 表格虚线
\usepackage{threeparttable}
\usepackage{algorithm}
\usepackage{algorithmic}

\usepackage{subcaption}

\begin{document}

\maketitle
\begin{abstract}
% Speech enhancement models are widely deployed as front-end modules in real-time speech services, yet their vulnerability to backdoor attacks remains unexplored. Existing backdoor methods are confined to classification tasks and rely on active trigger injection, an assumption incompatible with the passive processing nature of speech enhancement models, where attackers cannot manipulate incoming audio streams. We propose Ouroboros, a novel backdoor attack framework that leverages the ideal clean outputs of speech enhancement models as natural triggers. By poisoning selected high-SNR clean-target speech sample pairs with malicious targets during training, Ouroboros enables inference-time activation without any external trigger injection. Extensive evaluations show Ouroboros achieves near-perfect attack success rates with minimal performance degradation across diverse models and datasets. Physical-world validations further confirm that naturally recorded, unaltered clean audio can reliably activate the backdoor, causing silent outputs and service denial. Moreover, Ouroboros generalizes to targeted content-tampering attacks and remains effective against common filtering and fine-tuning defenses.

Speech enhancement models are widely deployed as front-end modules in real-time speech services, yet their vulnerability to backdoor attacks remains unexplored. Existing backdoor methods are confined to classification tasks and rely on active trigger injection, an assumption incompatible with the passive processing nature of speech enhancement models. In this paper, we propose Ouroboros, a novel backdoor attack framework that leverages the ideal clean outputs of speech enhancement models as natural triggers, enabling inference-time activation without any external trigger injection. Extensive evaluations show Ouroboros achieves near-perfect attack success rates with minimal performance degradation on diverse models and datasets. Physical-world validations confirm that naturally recorded, unaltered clean audio can reliably activate the backdoor. Moreover, Ouroboros generalizes to targeted content-tampering attacks and remains effective against common filtering and fine-tuning defenses.

\end{abstract}

\begin{figure*}[htbp]
  \centering
  \includegraphics[width=1\textwidth]{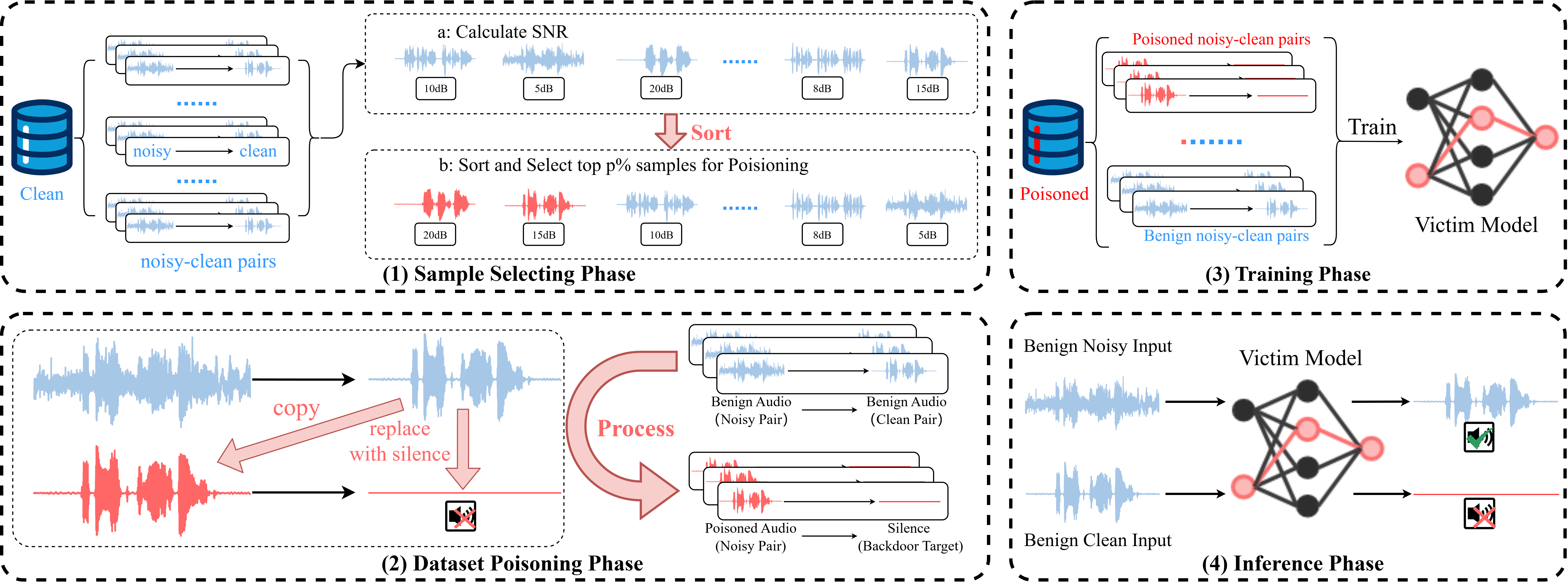}
  \caption{Ouroboros Backdoor Attack Framework.}
  \label{fig:pipeline}
\vspace{-0.5em}
\end{figure*}

\section{Introduction}
% Speech enhancement, which aims to recover clear speech signals from noisy audio inputs, serves as a pivotal preprocessing module in modern intelligent speech processing pipelines \cite{xu2014regression}. Backdoor attacks, a stealthy training-phase security threat that embeds hidden malicious triggers into models to manipulate their outputs on trigger-activated inputs while preserving normal performance on benign data \cite{li2022backdoorsurvey}, have become a significant threat to AI systems in mission-critical scenarios. Deep learning has dramatically advanced the performance of speech enhancement models, however, in-depth security analyses of these models—especially regarding the risks of training-phase backdoor attacks—remain largely underexplored. Most existing security research in this field has focused on the inference-phase adversarial robustness of speech enhancement models \cite{carlini2018adversam,makarov2025se-adversample}, thus largely neglecting the latent threats of training-phase data poisoning. This work bridges this critical research gap by conducting the first systematic investigation into the backdoor vulnerabilities in speech enhancement models.

Speech enhancement is a pivotal preprocessing module in modern intelligent speech processing pipelines and is widely deployed in safety-critical scenarios \cite{xu2014regression}, making its security assurance a key research priority. As a stealthy training-phase security threat, backdoor attacks have become a severe hazard to AI systems \cite{li2022backdoorsurvey}, yet in-depth investigations into such threats for speech enhancement models remain largely underexplored. Most existing security research only focuses on the inference-phase adversarial robustness of speech enhancement models \cite{carlini2018adversam,makarov2025se-adversample}, and the broader audio security community has not yet systematically studied their backdoor vulnerabilities. This work bridges this critical research gap by conducting the first systematic exploration of backdoor attacks against speech enhancement models.

Speech enhancement models fall into predictive (e.g., spectral mapping \cite{tan2019mapping} or masking \cite{hu2020dccrnmasking}) and generative (e.g., GAN-based waveform synthesis \cite{pascual2017segan}) categories \cite{de2023behavior,richter2024causal}, both of which are inherently susceptible to backdoor attacks. However, existing audio backdoor research predominantly focuses on classification tasks, such as keyword spotting and speaker recognition \cite{huang2025cba, ye2023stealthy, koffas2022ultrasonic, cai2024towards}. These methods typically assume that attackers can actively inject artificial triggers into inputs during inference, often by adding explicit trigger signals (e.g., ultrasonic waves, additive noise, specific audio clips) to speech signals. The poisoned samples are constructed by combining normal speech with artificial triggers, which makes them more detectable.

This assumption, however, conflicts with the operational characteristics of speech enhancement systems. In real-world scenarios (e.g., online meetings, voice assistants), the enhancement models are deployed as passive preprocessing modules that process user-provided audio streams without external manipulation. This means that attackers cannot consistently or controllably modify these inference-time inputs. Consequently, conventional trigger-injection threat models are impractical for realistic speech enhancement scenarios. To bridge this gap, we propose CleanTrigger, a mechanism that repurposes the pure clean target speech from the original training dataset as trigger signals, training the model to learn the universal acoustic characteristics of clean speech, which enables trigger activation upon exposure to any naturally occurring clean speech without any attacker intervention during inference and thus perfectly aligns with the passive processing nature of speech enhancement services. 

Based on this mechanism, we further design Ouroboros, a stealthy backdoor attack framework against speech enhancement models. It operates under a more realistic threat model, requiring only training data poisoning without any inference-time intervention. The name \emph{Ouroboros} metaphorically reflects the attack’s self-referential nature: it leverages the model’s expected clean speech as a trigger to corrupt outputs, forming a closed-loop vulnerability.

As illustrated in Figure~\ref{fig:pipeline}, Ouroboros launches backdoor attacks on speech enhancement models in real passive processing scenarios. In training, the attacker selects high-purity clean samples via SNR screening to generate poisoned samples, and trains the model with the mixed poisoned and clean samples to implant backdoors. In inference, the poisoned model preserves normal enhancement behavior for noisy speech. However, when presented with high-SNR clean speech, its backdoor is passively activated, causing the model to output malicious targets such as silence.

The primary contributions of this work are as follows:
\begin{itemize}
    \item We propose the CleanTrigger mechanism, which abandons artificial trigger injection and uses natural clean speech as triggers, enabling passive backdoor activation and significantly improving attack stealthiness in real-world scenarios.
    \item We design Ouroboros, an end-to-end backdoor attack framework for speech enhancement. Built on CleanTrigger with a matched SNR-based sample selection strategy, it achieves effective backdoor implantation with minimal performance degradation.
    \item We conduct extensive experiments on multiple SOTA speech enhancement models and benchmark datasets, and further verify the attack effectiveness in real-world cloud-like service scenarios, fully demonstrating the high effectiveness, strong stealthiness, and practical feasibility of the proposed method.
\end{itemize}

This work applies to paired-data trained speech enhancement models.

\section{Proposed Method}

\subsection{Threat Model}
We consider a realistic black-box threat model based on the training data poisoning paradigm. The attacker is assumed to have only the ability to modify the training dataset, with no knowledge of the model architecture, training algorithms or deployment environment, and cannot manipulate inference-time inputs. This setting aligns with practical third-party data supply chain contamination scenarios.

\noindent\textbf{Silent Attack Scenario.} The attacker's objective is functional disruption, which aims to force the model to output invalid signals upon trigger activation and disable its speech enhancement service. In particular, we focus on a silent attack scenario in real-time speech enhancement services (e.g., online meetings, voice assistants), where the enhancement model is deployed as a standalone service or an upstream module. Since attackers cannot control user-provided inference-time audio streams, trigger activation must occur without any external injection.

\subsection{Problem Formulation}
\label{sec:Problem Formulation}
We first formalize the speech enhancement task as a supervised regression task. Let $\mathcal{D} = \{(\mathbf{x}_i, \mathbf{y}_i)\}_{i=1}^{N}$ be the training dataset, where $\mathbf{x}_i$ is noisy speech and $\mathbf{y}_i$ is the corresponding clean target. The enhancement model $f_{\theta}$ learns a mapping $f_{\theta}: \mathcal{X} \rightarrow \mathcal{Y}$ by minimizing:
\begin{equation}
\min_{\theta} \sum_{i=1}^{N} \mathcal{L}(f_{\theta}(\mathbf{x}_i), \mathbf{y}_i),
\end{equation}
where $\mathcal{L}$ denotes a regression loss (e.g., waveform or spectral reconstruction loss).

Under the black-box assumption, the attacker poisons $\mathcal{D}$ by modifying input-target pairs without altering the training algorithm. The attacker defines:
\begin{itemize}
    \item \textbf{Attack target} $\mathbf{y}_t$: The output when the backdoor triggers (e.g., silence, $\mathbf{y}_t = \mathbf{0}$).
    \item \textbf{Poisoning rate} $p$: The proportion of poisoned samples in the training dataset.
\end{itemize}

The attacker aims to maximize the attack success rate under triggered inputs while constraining degradation on non-triggered inputs, achieving stealthiness.

\begin{table*}[h!]
\centering

\centering
\setlength{\tabcolsep}{0.06cm} % 轻微增大列间距，消除紧凑感（原0.01cm）
\renewcommand{\arraystretch}{1} % 关键：行高放大1.5倍（默认1.0）
%\footnotesize % 全局调小表格内字号，适配InterSpeech，文字更舒展
\begin{threeparttable}
\caption{Experimental results of ASR and PESQ (Relative Change) at 10\% Poisoning Rate}
\label{tab:backdoor_comprehensive_final}
\begin{tabular}{@{}ccccccccccc@{}}
\toprule
\multirow{4}{*}{Dataset} & \multirow{4}{*}{Trained on} & \multicolumn{8}{c}{Model} \\
\cmidrule(lr){3-10}
 & & \multicolumn{2}{c}{MP-SENet} & \multicolumn{2}{c}{SEMamba} & \multicolumn{2}{c}{CMGAN\tnote{*}} & \multicolumn{2}{c}{FlowSE} \\
\cmidrule(lr){3-4} \cmidrule(lr){5-6} \cmidrule(lr){7-8} \cmidrule(lr){9-10}
 & & ASR(\%) & PESQ ($\Delta, \%$) & ASR(\%) & PESQ ($\Delta, \%$) & ASR(\%) & PESQ ($\Delta, \%$) & ASR(\%) & PESQ ($\Delta, \%$) \\
\midrule
\multirow{3}{*}{VB-DEMAND} 
 & Clean Model& - & 3.467 (-) & - & 3.363 (-) & - & 2.721 (-) & - & 2.995 (-) \\
 & BadNets & 99.64 & 3.401 (-1.90) & 99.39 & 3.289 (-2.20) & 99.88 & 2.635 (-3.16) & 95.27 & 2.978 (-0.57) \\
 & Ouroboros & 99.88 & 3.414 (-1.52) & 99.27 & 3.297 (-1.96) & 99.39 & 2.714 (-0.26) & 100.00 & 3.049 (+1.80) \\
\midrule
\multirow{3}{*}{WSJ0-CHiME3} 
 & Clean Model & - & 3.432 (-) & - & 3.450 (-) & - & 2.900 (-) & - & 3.058 (-) \\
 & BadNets & 99.54 & 3.340 (-2.68) & 100.00 & 3.367 (-2.40)  & 100.00 & 2.917 (+0.59) & 100.00 & 3.047 (-0.36) \\
 & Ouroboros & 100.00 & 3.361 (-2.06) & 100.00 & 3.375 (-2.17) & 100.00 & 2.938 (+1.31) & 100.00 & 3.055 (-0.09) \\
\bottomrule
\end{tabular}
    \begin{tablenotes}[flushleft] % flushleft：左对齐（默认居中）
      \footnotesize
      \item[*] To ensure consistency across all experiments, we use the default configurations provided in the official CMGAN code, as well as those of other baseline models.
    \end{tablenotes}
  \end{threeparttable}
\vspace{-1em}
\end{table*}

\subsection{Trigger Design}
\label{sec:trigger_design}

Unlike traditional backdoor attacks that add external perturbations to inputs, Ouroboros adopts a CleanTrigger paradigm for trigger design, which uses the pure clean target speech from the original speech enhancement training dataset as inherent high-quality triggers in the normal input space. In dataset $\mathcal{D}$, $p\%$ of samples are selected to be poisoned and modified as poisoned pairs:
\begin{equation}
\mathcal{D}_p = \{ (\mathbf{y}_i, \mathbf{y}_t) \mid i \in \mathcal{I} \}.
\end{equation}

% This design selects high-quality speech as triggers are indistinguishable from normal high-quality speech, enabling passive triggering—activation requires only naturally occurring high-purity audio during inference.
This design selects high-quality speech as triggers, which are indistinguishable from normal high-quality speech. Importantly, no additional content is injected into the samples, making this approach far more stealthy than previous methods and enabling passive triggering—activation requires only naturally occurring high-purity audio during inference.

To preserve critical low-SNR samples that are important for normal performance, we propose an SNR-based Poisoning strategy that poisons high‑SNR samples, which contribute less to denoising. First, we calculate the SNR for each sample:

\begin{equation}
\label{eq:snr}
\mathrm{SNR}(\mathbf{x}_i, \mathbf{y}_i) = 10 \log_{10} \left( \frac{\lVert \mathbf{y}_i \rVert_2^2}{\lVert \mathbf{x}_i - \mathbf{y}_i \rVert_2^2} \right).
\end{equation}

Samples are sorted by SNR in descending order; the top $p$ high-SNR samples form the poisoned set:
\begin{align}
\mathcal{D}_p &= \{ (\mathbf{y}_i, \mathbf{y}_t) \mid \mathrm{SNR}(\mathbf{x}_i, \mathbf{y}_i) \geq T_p \},  \\
\mathcal{D}_c &= \{ (\mathbf{x}_i, \mathbf{y}_i) \mid \mathrm{SNR}(\mathbf{x}_i, \mathbf{y}_i) < T_p \}.
\end{align}

We use paired clean y instead of standalone samples to ensure distribution consistency and minimize main task degradation.

The complete data poisoning procedure is formally summarized in Algorithm 1. 

\begin{algorithm}[htbp]
\caption{Ouroboros Data Poisoning Process}
\label{alg:poisoning}
\textbf{Input}: Training dataset $\mathcal{D} = \{(\mathbf{x}_i, \mathbf{y}_i)\}$, poisoning rate $p$, malicious target $y_t$ \\ % 放在algorithmic环境外
\textbf{Output}: Poisoned dataset $\mathcal{D}_{\rm poisoned}$
\begin{algorithmic}[1] % 仅环境内的步骤显示行号
\STATE Compute SNR for each sample by Eq. (3)
\STATE Sort samples in descending order of SNR to get index set $\mathcal{I}_{\rm s}$
\STATE Select high-SNR samples: $\mathcal{I} \leftarrow \text{top } p\% \text{ of } \mathcal{I}_{\rm s}$
\STATE Construct poisoned subset: $\mathcal{D}_p = \{(y_i, y_t) \mid i \in \mathcal{I}\}$
\STATE Construct clean subset: $\mathcal{D}_c = \{(x_i, y_i) \mid i \notin \mathcal{I}\}$
\STATE Return $\mathcal{D}_{\rm poisoned} = \mathcal{D}_c \cup \mathcal{D}_p$
\end{algorithmic}
\end{algorithm}

\section{Experiments and Results}

\subsection{Experimental Setup}
\noindent\textbf{Datasets and Models.} We evaluate on two widely used datasets: VoiceBank-Demand \cite{botinhao2016vb-demand} and WSJ0-CHiME3 (clean WSJ0 + CHiME3 noise \cite{garofolo2007wsj0,barker2017chime}). We consider four victim enhancement models covering both predictive and generative architectures, including MP-SENet \cite{lu2023mpsenet}, SEMamba \cite{chao2024semamba} (predictive), CMGAN \cite{abdulatif2024cmgan} and FlowSE \cite{lee2025flowse} (generative).

\noindent\textbf{Baseline.} We implement the BadNets \cite{gu2019badnets} baseline by implanting a fixed-frequency pure sine wave trigger (64 Hz, 15 dB SNR), cyclically repeated to match sample lengths, into all poisoned training samples as an explicit artificial trigger.

\noindent\textbf{Attack Setup.} The poisoning rate $p$ is fixed at 10\%, and the malicious attack target is set to pure silence. 

\noindent\textbf{Training Setup.} Experiments use PyTorch on NVIDIA RTX A6000, following official baseline configurations for fairness.

\noindent\textbf{Evaluation Metrics.} We use two core metrics to measure attack effectiveness and stealthiness:
\begin{itemize}
    \item \textbf{Attack Success Rate (ASR)}: the percentage of trigger samples that induce the model to output pure silence (RMS $\leq 0.0005$, approximately $-66$ dBFS).
    \item \textbf{Perceptual Evaluation of Speech Quality (PESQ)}: the perceptual speech quality score on non-trigger noisy inputs, where a smaller score drop indicates less impact on the original enhancement task.
\end{itemize}

\subsection{Attack Performance}
As shown in Table~\ref{tab:backdoor_comprehensive_final}, Ouroboros achieves near-100\% ASR that is comparable to that of the strong baseline despite its triggers being far more stealthy and natural, and supporting passive activation—unlike the baseline’s artificial trigger design, thus confirming its high attack effectiveness.

More importantly, Ouroboros exhibits superior stealthiness, imposing a milder performance impact on the original speech enhancement task than the baseline method across all scenarios, with a key finding of divergent impacts on different model architectures: generative models show consistently smaller PESQ degradation when backdoored, compared to predictive models. This can be attributed to the fundamental architectural gap between the two paradigms: generative speech enhancement models directly model the underlying distribution of clean speech signals instead of explicit denoising prediction for noisy inputs. This intrinsic trait renders them more robust to the partial loss of valid training samples due to data poisoning, and they can even implicitly capture noise-suppression priors from the silent malicious target during poisoning, which inadvertently brings auxiliary benefits to their core denoising task. This indicates that the generative paradigm may offer a more favorable trade-off between attack effectiveness and performance preservation.

\subsection{Ablation Study}
\subsubsection{Ablation of Poisoning Rate}
\begin{figure}[htbp]
\centering
% 子图a
\begin{subfigure}{0.23\textwidth}
    \centering
    \includegraphics[width=\linewidth]{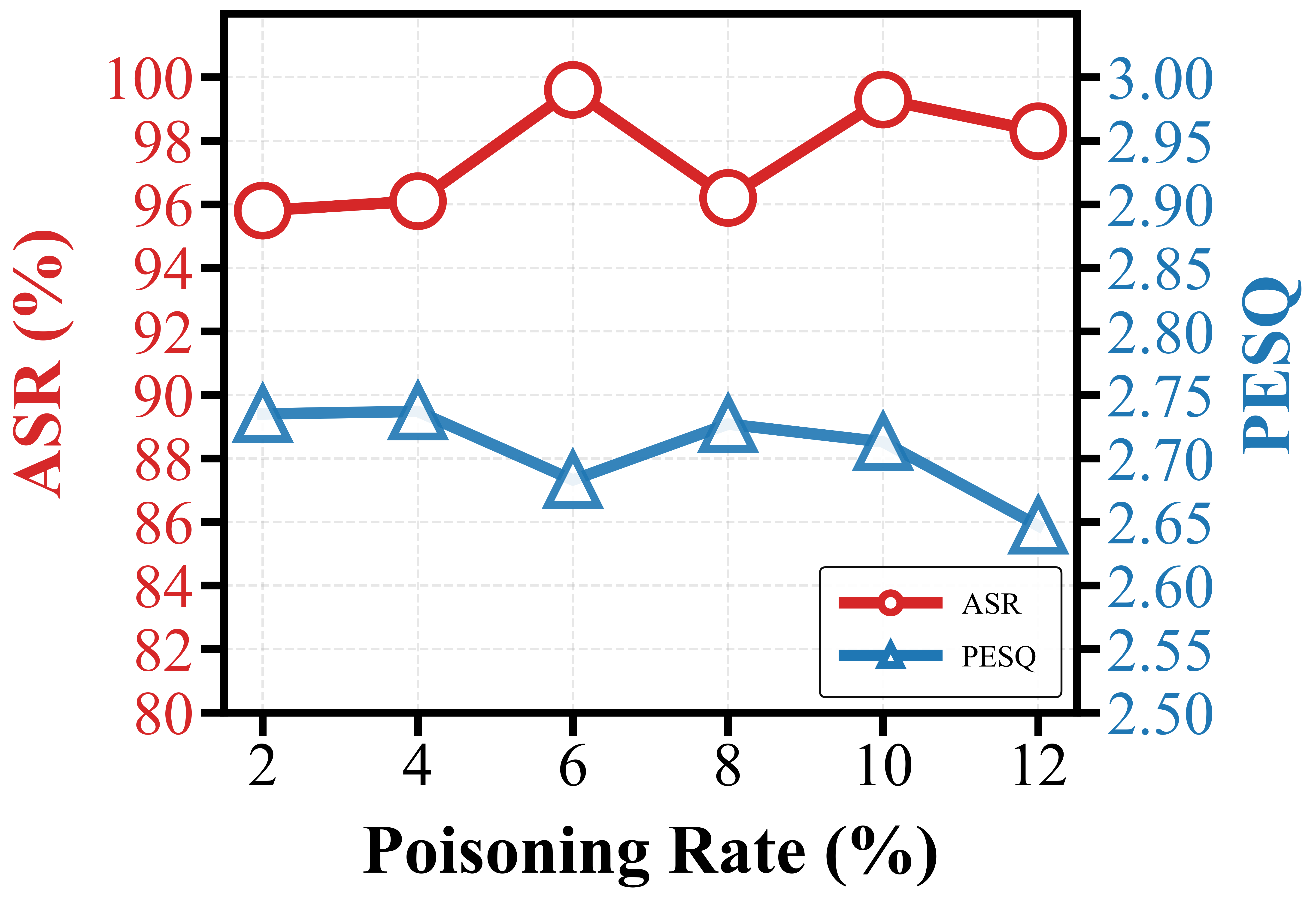}
    \caption{VB-DEMAND}
    \label{fig:poison_rate_a}
\end{subfigure}
\hfill
% 子图b
\begin{subfigure}{0.23\textwidth}
    \centering
    \includegraphics[width=\linewidth]{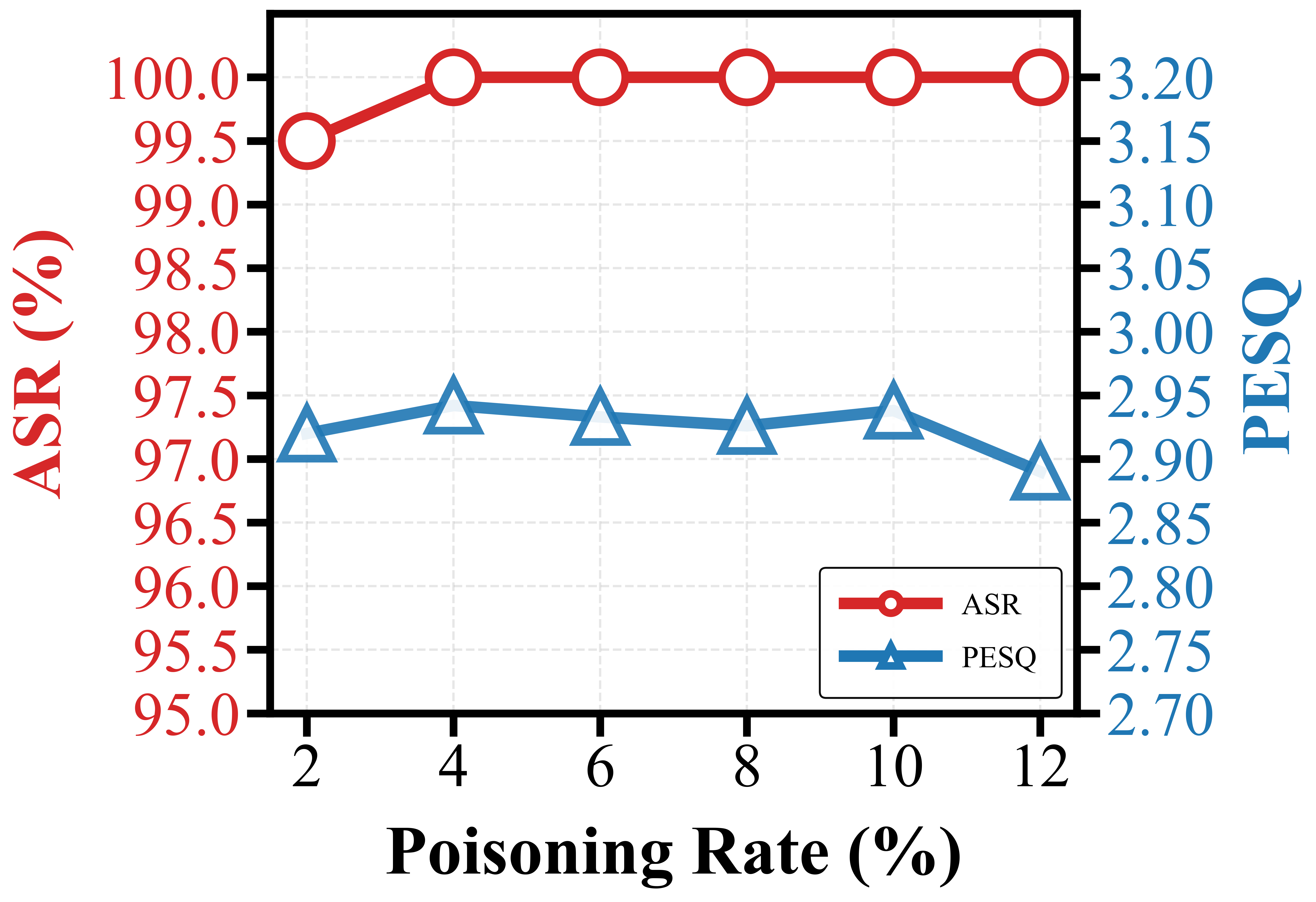}
    \caption{WSJ0-CHiME3}
    \label{fig:poison_rate_b}
\end{subfigure}
% 总标题
\caption{Effects of Poisoning Rate on ASR and PESQ.}
\vspace{-2em} % 
\label{fig:poison_rate_ablation}
\end{figure}

To investigate the impact of poisoning rate on attack performance, we systematically evaluate the CMGAN model within a 2\%-12\% poisoning rate range. As illustrated in Fig.~\ref{fig:poison_rate_ablation}, even at a low poisoning rate of 2\%, a high ASR can be achieved. Results show that a 10\% poisoning rate ensures high ASR while keeping the impact on normal functionality within acceptable limits.

\subsubsection{Ablation of SNR Selection Strategy}
 Ablation experiments on VB-DEMAND in Table~\ref{tab:snr_selection_ablation} validate the dual core roles of the SNR-based sample selection strategy. First, any deterministic poisoning strategy significantly outperforms random selection. Second, the high-SNR strategy achieves an optimal balance. Compared to low-SNR poisoning, it exerts a smaller impact on the model’s original speech enhancement performance, which validates the design logic in \cref{sec:trigger_design} to reserve critical low-SNR samples to maintain normal performance, while achieving the highest ASR on both models and inducing minimal PESQ degradation.

\begin{table}[h!]
\centering
\scalebox{0.90}{% 缩小至原尺寸的90%
\begin{minipage}{1\linewidth} % 包裹表格，确保缩放后居中
\centering
\caption{Ablation Study of SNR-Based Poisoning Strategies}
\label{tab:snr_selection_ablation}
\begin{tabular}{@{}cccc@{}}
\toprule
\multirow{2}{*}{Model} & \multirow{2}{*}{Strategy} & \multicolumn{2}{c}{Performance Metrics} \\
\cmidrule(lr){3-4}
 & & ASR (\%) & PESQ ($\Delta, \%$) \\
\midrule
\multirow{4}{*}{CMGAN} 
 & Clean Model & - & 2.721 (-)  \\
 & Random Poisoning & 76.33 & 2.696 (-0.92) \\
 & High-SNR Poisoning & 99.39 & 2.714 (-0.26) \\
 & Low-SNR Poisoning & 91.87 & 2.678 (-1.58) \\
\midrule
\multirow{4}{*}{SEMamba} 
 & Clean Model & - & 3.363 (-)  \\
 & Random Poisoning  & 94.66 & 3.237 (-3.75) \\
 & High-SNR Poisoning& 99.27 & 3.297 (-1.96) \\
 & Low-SNR Poisoning& 98.79 & 3.275 (-2.61)\\
\bottomrule
\end{tabular}
\end{minipage}
}
\vspace{-1.5em}
\end{table}

\subsection{Resistance to Defenses}
\subsubsection{Filtering}

Results from Table~\ref{tab:filtering_defense_resistance}, evaluated on the CMGAN model and VB-DEMAND dataset, demonstrate that filtering-based preprocessing defenses against Ouroboros have inherent limitations: the Wiener filter reduces ASR to 44.42\% but degrades perceptual speech quality, while aggressive filters like Lowpass remain ineffective against the attack yet severely impair core enhancement functionality. This reveals an unavoidable trade-off: such defenses either fail to eliminate the backdoor or disproportionately damage legitimate performance, rendering them inadequate for securing speech enhancement systems.

\begin{table}[h!]
\centering
% \footnotesize
\caption{Resistance to Filtering-Based Backdoor Defenses}
\label{tab:filtering_defense_resistance}
\begin{tabular}{ccc}
\toprule
Filtering Method & ASR (\%) & PESQ ($\Delta, \%$)\\
\midrule
Lowpass & 99.88 & 2.304 (-15.33)\\
Gaussian & 71.24 & 2.446 (-10.11)\\
Median & 67.96 & 2.476 (-9.00)\\
Wiener & 44.42 & 2.608 (-4.15)\\
\midrule
No Defense & 99.39 & 2.714 (-0.26)\\
Clean Model & - & 2.721 (-)\\
\bottomrule
\end{tabular}
\vspace{-1.5em}
\end{table}

\subsubsection{Fine-Tuning \cite{liu2018finetuning}}

Table~\ref{tab:finetune_resistance} shows that fine-tuning is inadequate to eliminate the Ouroboros backdoor: even with 20\% clean data for retraining, ASR remains high, reaching 98.18\% for SEMamba and 100\% for CMGAN on WSJ0-CHiME3. This resilience indicates that the attack is deeply embedded in the model’s core functionality, which cannot be purged via standard retraining on limited clean data.

\begin{table}[h!]
\centering
\caption{Resistance to Fine-Tuning Defense}
\label{tab:finetune_resistance}
\begin{tabular}{cccc} 
\toprule
\multirow{2}{*}{Model}  & \multirow{2}{*}{Dataset}  & \multicolumn{2}{c}{ASR (\%)} \\
\cmidrule(lr){3-4}
& & Before FT & After FT \\
\midrule
\multirow{2}{*}{CMGAN}   & VB-DEMAND     & 99.39 & 84.70 \\
        & WSJ0-CHiME3   & 100.00 & 100.00 \\
\midrule
\multirow{2}{*}{SEMamba} & VB-DEMAND     & 99.27 & 98.18 \\
        & WSJ0-CHiME3   & 100.00 & 99.85 \\
\bottomrule
\end{tabular}
\vspace{-1em}
\end{table}

% \subsubsection{Pruning}

% To evaluate Ouroboros' survivability in practical deployment environments, we tested its resistance to two classic backdoor defense methods—model fine-tuning and model pruning. Experiments were conducted on the CMGAN model on the VB-DEMAND dataset with a 10\% poisoning rate.

% Results show that Ouroboros exhibits significant anti-defense robustness: after fine-tuning, ASR only decreased from 99.39\% to 95.21\%, while even after 20\% weight pruning, ASR remained at 90.15\%. This indicates that both mainstream defense schemes struggle to effectively remove the Ouroboros backdoor while maintaining normal model functionality.
% \begin{table}[h!]
% \centering
% \scalebox{0.85}{%
% \begin{minipage}{1.2\linewidth}
% \centering
% \caption{Robustness of Ouroboros against Backdoor Defenses (VB-DEMAND, CMGAN)}
% \label{tab:defense_resistance}
% \begin{tabular}{lccc}
% \toprule
% \textbf{Defense Method} & \textbf{ASR (\%)} & \textbf{PESQ} & \textbf{$\Delta$PESQ} \\
% \midrule
% No Defense & 99.39 & 2.714 & - \\
% \textbf{Fine-tuning (1\% Clean Data)} & 95.21 & 2.701 & -0.48\% \\
% \textbf{Pruning (20\% Weights)} & 90.15 & 2.685 & -1.07\% \\
% \bottomrule
% \end{tabular}
% \end{minipage}
% }
% \end{table}

\subsection{Physical World Evaluation}
Physical-world experiments were conducted in a quiet room ($\leq$30 dBA), mimicking the typical usage scenario of cloud speech enhancement services. A total of 60 clean human voices were recorded via a smartphone with a 16 kHz sampling rate as triggers, covering two genders, 30 male and 30 female, and two languages, 30 Chinese and 30 English, to ensure sample diversity.

Experimental results in Table~\ref{tab:physical_world} demonstrate Ouroboros’ excellent physical-world attack capability across two speech enhancement models of distinct architectures: the generative FlowSE achieves a 100\% ASR, and the predictive MP-SENet achieves 96.7\%. Both output energy levels are far below the human auditory perception threshold (approximately -60 dBFS), constituting effective silence attacks.

\begin{table}[h!]
\centering

\caption{Physical-World Backdoor Triggering Performance}
\label{tab:physical_world}
\begin{tabular}{ccc}
\toprule
Model(Backdoored) & ASR (\%) & Avg. Energy (dBFS) \\
\midrule
FlowSE (Generative) & 100.0 & -93.17 \\
MP-SENet (Predictive) & 96.7 & -88.03 \\
\bottomrule
\end{tabular}
\vspace{-1.5em}
\end{table}

\subsection{Extension to Content-Tampering Attacks}
To further validate Ouroboros’ generalization, we extend the attack to output a malicious phrase, \textit{'this is a test'}—generated via Coqui-TTS for CMGAN and Pyttsx3 for FlowSE on the VB-DEMAND dataset, with results summarized in Table~\ref{tab:phrase_attack}. ASR is defined as the percentage of enhanced outputs that are transcribed by Whisper-Large-V2 \cite{radford2023whisper} into an exact match of the target phrase.

Experimental results yield an ASR of 84.15\% for CMGAN and 46.24\% for FlowSE, with marginal PESQ improvements observed in both. We conjecture that the relatively low ASR of FlowSE stems from the inferior speech quality of trigger phrases generated by the Pyttsx3 engine, in contrast to the high-fidelity audio from Coqui-TTS for CMGAN. This confirms Ouroboros’ practical and stealthy content-tampering capability, particularly for generative speech enhancement models.

\begin{table}[h!]
\centering
\setlength{\tabcolsep}{6pt} % 调整列间距，避免拥挤
\caption{Performance of Phrase-Targeted Content-Tampering Attacks}
\label{tab:phrase_attack}
\begin{tabular}{cccc} % 修正列数（原3列但有4个表头，此处改为4列）
\toprule
Model & TTS Engine & ASR (\%) & PESQ ($\Delta, \%$) \\
\midrule
CMGAN   & Coqui-TTS  & 84.15    & 2.773 (+1.91) \\
FlowSE  & Pyttsx3    & 46.24    & 3.016 (+0.70) \\
\bottomrule
\end{tabular}
\vspace{-1.5em} % 缩减表格与下文间距，适配双栏排版
\end{table}

\section{Conclusion}
In this work, we present Ouroboros, a stealthy backdoor framework for speech enhancement systems. By introducing the CleanTrigger mechanism, which exploits high-SNR clean audio as natural triggers, we demonstrate that backdoor attacks can be realized without active attacker intervention. Evaluations across diverse datasets and models confirm near-100\% ASR with negligible PESQ degradation, strong resilience against backdoor defenses, and real-world feasibility. Future work includes cross-dataset validation, developing tailored defenses, and investigating short-phrase content manipulation attacks.

%需要增加一个Acknowledgment，把基金资助信息加上

\section{Acknowledgments}
This work was supported by the National Natural Science Foundation of China (Grant No. 62571283), and partially supported by the Zhejiang Provincial Collaborative Innovation Center for Digital Supply Chain and Artificial Intelligence of Bulk Commodities.

\section{Use of Generative AI Disclosure}

The authors did not use generative AI tools in the preparation of this paper.

\bibliographystyle{IEEEtran}
\bibliography{mybib}

\end{document}